\documentclass[preprint,aps]{revtex4}
\usepackage{graphics}
\usepackage{graphicx}

\newcommand{\BV}{\left(\begin{array}{c}}
\newcommand{\EV}{\end{array}\right)}
\newcommand{\BM}{\left(\begin{array}{cc}}

\newcommand{\beqry}{\begin{eqnarray}}
\newcommand{\eeqry}{\end{eqnarray}}

\begin{document}

\title{Order from disorder in closed systems via time reversal violation}
\author{T.\ Goldman and D.H.\ Sharp \\
%\affiliation
Theoretical Division \\
Los Alamos National Laboratory \\
Los Alamos, NM 87545 USA}

\vspace*{-0.75in}
\begin{flushright}
LA-UR-11-11377
\end{flushright}
\vspace*{0.75in}

\begin{abstract}

Definitions of entropy usually assume time-reversal (T) invariance of interactions,
yet microscopically T is known to be violated. We present a detailed computational
example of (uncharged) particle species separation (Maxwell demon) using an 
interaction that violates both parity (P) and T so that PT is preserved, consistent 
with the CPT invariance required in quantum field theory (C is charge conjugation). 
This illustrates how T-violating forces can produce more organized states from 
disorganized ones, contrary to expectations based on increase of entropy. We 
also outline several scenarios in which T-violating forces could lead to an 
organized state in the early Universe, starting from a still earlier disorganized 
state. 

\end{abstract}

\maketitle

PACS: {11.30.Er, 98.80.-k,  98.80.Cq} 

\section{Introduction}

A central problem of statistical mechanics has been to understand the monotonic
increase of thermodynamic entropy in the presence of time-reversal-invariant
dynamics. The resolution of this problem is based on the observation that the
number of available states with higher entropy, for all but the simplest systems,
far exceeds those accessible with the same or lower entropy. An organized state
corresponds to one in which a small subset of the available states is preferentially
occupied. The result is a probabilistic conclusion that on average entropy increases
uniformly, although local fluctuations are possible in principle. (An instructive
example from everyday experience has been given by T.D.\ Lee\cite{tdl}.)

Here we discuss a related question: If the dynamics violates time reversal
invariance, is it still true that entropy must increase? Or, with the defined
arrow of time, is it possible that entropy must in fact decrease, at least 
for sufficiently large T-violation? 

If true, the latter possibility raises some intriguing questions when applied
to the early Universe: If conventionally-defined entropy can decrease when
T-violation is sufficiently large, there appear to be mechanisms that could
lead to the minimal entropy state assumed in the Big Bang hypothesis, even
if the universe was in a high entropy (disorganized) state at times prior to
the Big Bang. Several such mechanisms are discussed here, along with their 
bearing on such matters as baryon asymmetry. 

\section{Example 1: Scattering of Distinguishable Particles in 1 Dimension}

The example we discuss here is from classical physics, but is chosen so as
to be consistent with the CPT theorem \cite{cpt}, which is one of the basic
constraints of conventional quantum field theory. The CPT theorem states that
the combination of charge conjugation, parity inversion and time reversal
operations leaves the world Lagrangian invariant.  

Consider two distinguishable types of classical particle A and B  moving in a
1 + 1 dimensional manifold. If the particles are uncharged, the overall Lorentz
symmetry constraint in a flat space-time requires that the combination of parity
(P) and time-reversal (T) is an invariance of the dynamics. Combined PT invariance 
is necessary to satisfy the requirement that all amplitudes be restricted to the proper, 
orthochronous sector of the Lorentz transformations; P and T need not separately 
be invariances of the dynamics. Access to a wide range of papers that consider P 
and T violation while preserving PT can be found at $<http://ptsymmetry.net>$. 
(The CPT theorem has not been demonstrated in curved space-times.\cite{kost} ) 

We now introduce an interaction which violates P and T separately, but
preserves the product. Specifically, we suppose the probability is unity that A, 
incoming from the left, and B, incoming from the right, reaches the final state 
in which A continues to the right and B continues to the left. (This 
scattering and all others described below are taken to be perfectly elastic, 
energy-momentum conserving scatterings. For convenience in our 
calculations, we will also assume equal masses for the distinguishable 
particles, so that energy-momentum obtains trivially.)  Further, we assume 
that the amplitude for the other possible final state (A reflected back to the 
left and B back to the right) vanishes identically. That is, 
\begin{eqnarray} 
P(p_A > 0, p_B < 0 \rightarrow p'_A =p_A > 0, p'_B = p_B < 0) & = & 1 \nonumber \\
P(p_A > 0, p_B < 0 \rightarrow p'_A= p_B < 0, p'_B =p_A> 0) & = & 0 \label{maxtv1}
\end{eqnarray} 
where the values of the momenta are unchanged in the first case, and 
the values for particles A and B are exchanged in the second. 

However, for the parity reversed initial state ( A incoming from the right and B 
incoming from the left), we suppose that the probability is unity  for scattering to 
the final state in which A reflects back to the right and B reflects back to the left 
and that there is zero probability for each particle to continue forward. That is, 
\begin{eqnarray} 
P(p_A < 0, p_B > 0 \rightarrow p'_A = p_B > 0, p'_B = p_A < 0) & = & 1 \nonumber \\
P(p_A < 0, p_B > 0 \rightarrow p'_A = p_A < 0, p'_B = p_B > 0) & = & 0 \label{maxtv2}
\end{eqnarray} 
These probabilities clearly maximally violate P and T
separately. On applying both transformations, it is also apparent that
the PT product is nonetheless an invariant; we take all of the AA and BB forward 
and backward scattering probabilities to be equal. 

We observe that the equilibrium state of this dynamics is a separation of
any initial mixture of A and B into a segregated system with only particles
of type A to the right and only particles of type B to the left, where the
location of the dividing line is determined by the relative numbers of the
two particle types. The dynamics has acted as a Maxwell demon\cite{MR} to
produce a state which has lower entropy (as defined conventionally) than the
normally expected one of maximal mixing of the two particle types. 

Note, though, that no work can be extracted from the separated system
because the particles will not tend to remix unless the dynamics changes. 
Preservation of the concept that entropy can only increase now clearly  
requires  some different definition of entropy in order to assign the
separated system a higher value of the entropy than the fully mixed one.
That is, the final state is achieved not due to directed energy flow or
externally imposed
chemical potential differences, but is an equilibrium different from the
one expected with P- and T-conserving dynamics. It is the fully mixed
initial state that is out of equilibrium! 

We have verified the conclusions above in this 1+1-dimensional case 
by means of Monte Carlo simulation programs.  

We first wrote a Monte Carlo simulation program with T-conserving 
dynamics, where all of the probabilities in Eqs.(\ref{maxtv1},\ref{maxtv2}) 
are equal. The program populates the phase
space with particles A and B having equal temperatures and uniformly mixed
initial spatial distributions. Scattering is introduced when any two particles
propagate into close proximity. We have verified that with the P,T-conserving 
dynamics the system maintains the expected equilibrium distribution and
the two types of particles remain fully mixed, although separation fluctuations
of one standard deviation either way quite clearly occur. Following the motion
of a few particles chosen at random shows that they do propagate and scatter
as expected. Figure 1 displays the uniform spatial mixing (up to fluctuations),
temperature equality and phase space distribution in the presence of scattering
but in the absence of a T-violating interaction.

\begin{figure} 
\includegraphics[width=122mm]
{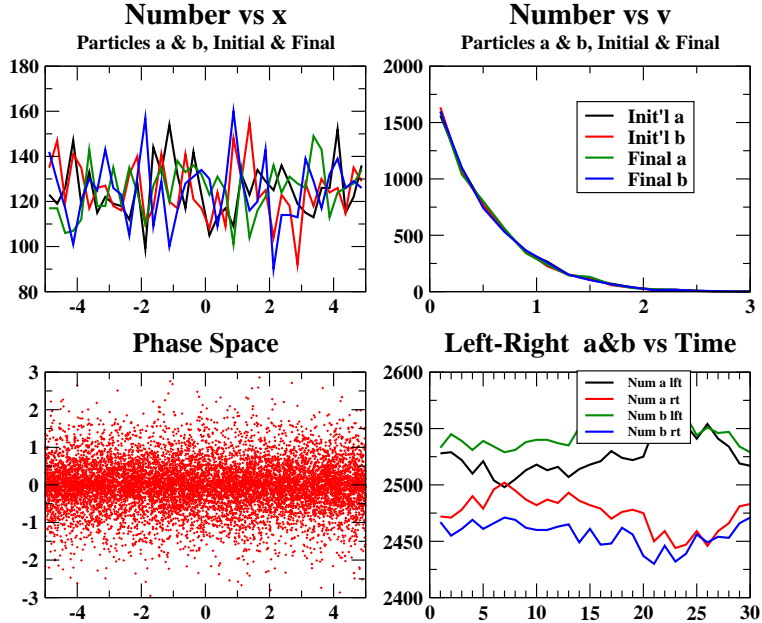}
\caption{(Color online) T-conserving distributions and time evolution. a) Upper left 
panel: Binned initial and final numbers of particles vs. position along line; color 
coding is the same as in the upper right panel b) which shows the velocity 
distributions. The phase space distribution, ignoring particle type, is shown 
in the lower left panel, c). The time variation of the number of particles of each 
type to the left or right of the center line is shown in the lower right panel, d). }
\label{fig:tconsv}
\end{figure}

\begin{figure}
\includegraphics[width=122mm]
{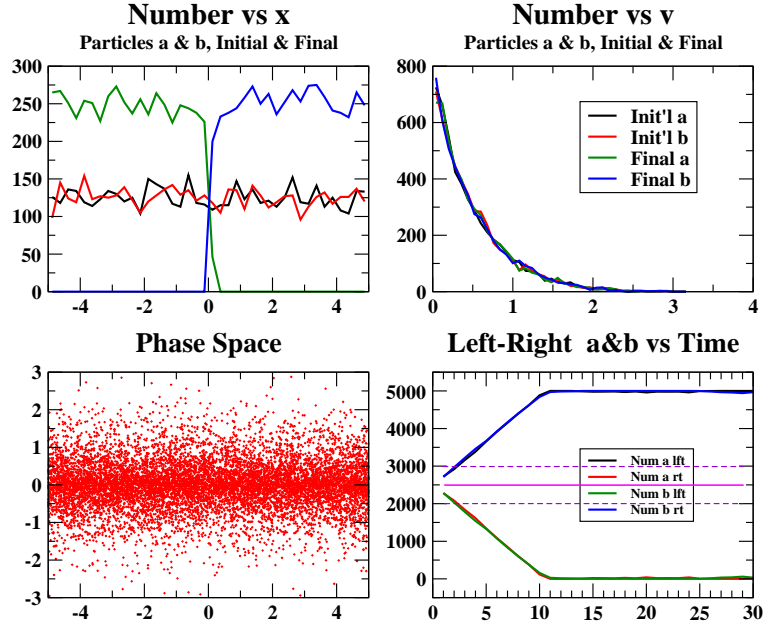}
\caption{(Color online) T-violating distributions and time evolution with 
maximal violation. Panels and color coding as in Fig.1; note different scale 
in phase space distribution. Upper left panel a) shows initial distributions fully 
mixed and final distributions well separated except for small central transition 
region. The dashed lines in the lower right panel d) show the one standard 
deviation expectation from the number of each species of particle.}
\label{fig:tviolmax}
\end{figure}

\begin{figure}
\includegraphics[width=122mm]
{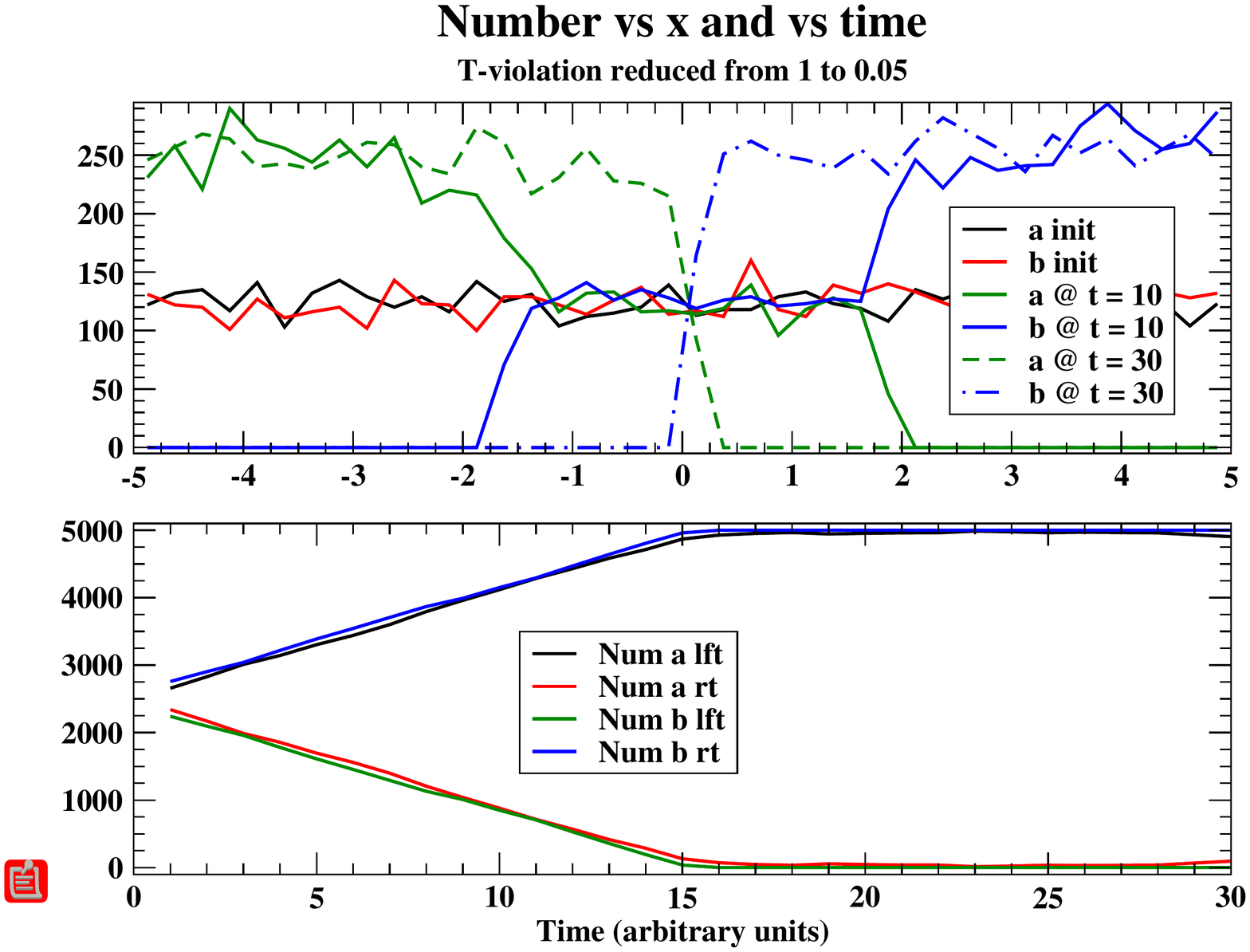}
\caption{(Color online) T-violating distributions a) and time evolution b) with 
reduced violation (5\% of maximal).  Note the longer time scale for the
system to separate vs. that in Fig.2.}
\label{fig:tviolless}
\end{figure}

We then changed the dynamics so as to satisfy the T-violating conditions 
described above in Eqs.(\ref{maxtv1},\ref{maxtv2}): 
we require that if A has negative velocity and B has a
positive value, no scattering takes place, but if A has positive velocity
and B has negative value, then the particles reflect (in the easiest to
understand, equal mass case, satisfying energy-momentum 
conservation) with A acquiring the negative velocity of B and B the
positive one of A. In this case, the initial uniformly mixed distribution
undergoes the expected segregation, with particles of type B flowing to
positive spatial locations and particles of type A flowing to negative
ones. The separation occurs quite rapidly for maximal violation. Figure
2 displays the initial uniform spatial mixing (up to fluctuations) and
the clear separation of A and B species after the T-violating interactions
have been turned on to affect the scattering. Temperature equality is
maintained and the phase space distribution, ignoring the particle type,
is qualitatively unaltered; the numerical evidence indicates that the 
phase space volume is conserved. The final state equilibrium is attained 
after approximately 10 time units. 

Our results also show that if we reduce the effective strength of P and 
T-violation by limiting the (random) fraction of AB scatters that violate
P and T to 5\% 
	(i.e., the zeros on the rhs of Eqs.(\ref{maxtv1},\ref{maxtv2}) are replaced 
	by 0.95)
and require the remaining fraction to satisfy these invariances, 
the same segregation develops. This is shown in Figure 3. (Temperature
and phase space distributions are unaffected and so are not repeated.) 
Note that, as expected, it takes longer for the final state equilibrium to be 
reached; this is now attained after approximately 15 time units.

The fractional amount of separation in the population increases linearly with
time in both cases. Examining the separation before it goes to completion 
shows an interesting pattern: The extreme positive and negative ends 
of the line become saturated with particles B and A respectively, but an 
intermediate, quite mixed, region remains. (Note that this is different 
from the conventional Maxwell demon picture.)  Observation of the 
track of randomly chosen particles confirms this: An A particle that 
happens to originate in a negative region tends to remain there, while 
one that originates in a positive region moves, albeit with unsteady 
fluctuations, towards the negative region, and vice versa for B particles. 
This directly displays the feature that for A-B scattering, the built-in 
T-violation of the scattering matrix causes A particles with positive 
velocity to be reflected but allows those with negative velocity to 
proceed unhindered, and conversely for the B particles. 
A similar phenomenon (termed "rectification") has been observed by 
Wan et al. in a numerical simulation of bacterial motion in the presence 
of barriers\cite{CR} that reproduces experimental observations of the 
same system. 

\section{Example 2: Scattering of Particles in 3 Dimensions}

\begin{figure}
\includegraphics[width=122mm]
{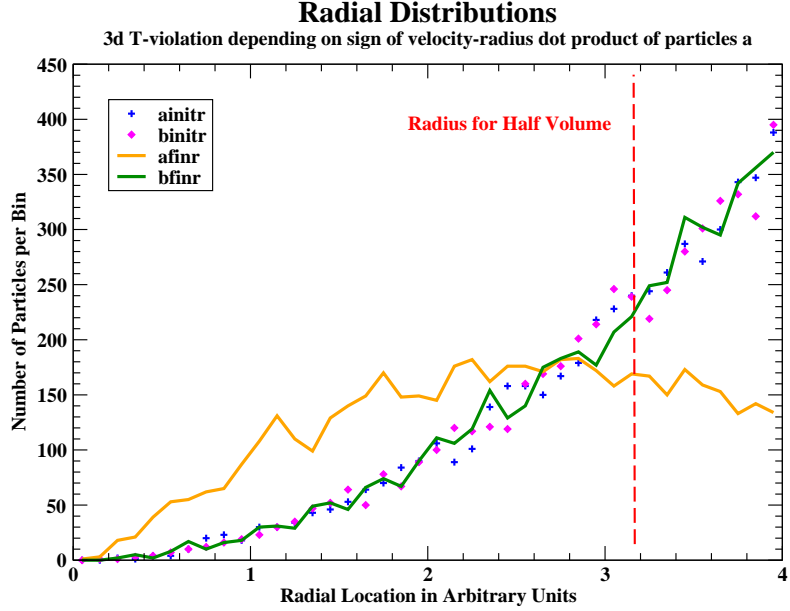}
\caption{(Color online) T-violating distributions when the scattering depends 
on  the sign of the dot product of the velocity of one species with its radial vector.
The initial radial distributions (ainitr, binitr) are constructed to be uniform. After 
some time, the species (a) subject to the T-violating scattering has been significantly 
compacted to the inner half-volume of the reference sphere (afinr) of radius 4 units. }
\label{fig:tviolra}
\end{figure}

\begin{figure}
\includegraphics[width=122mm]
{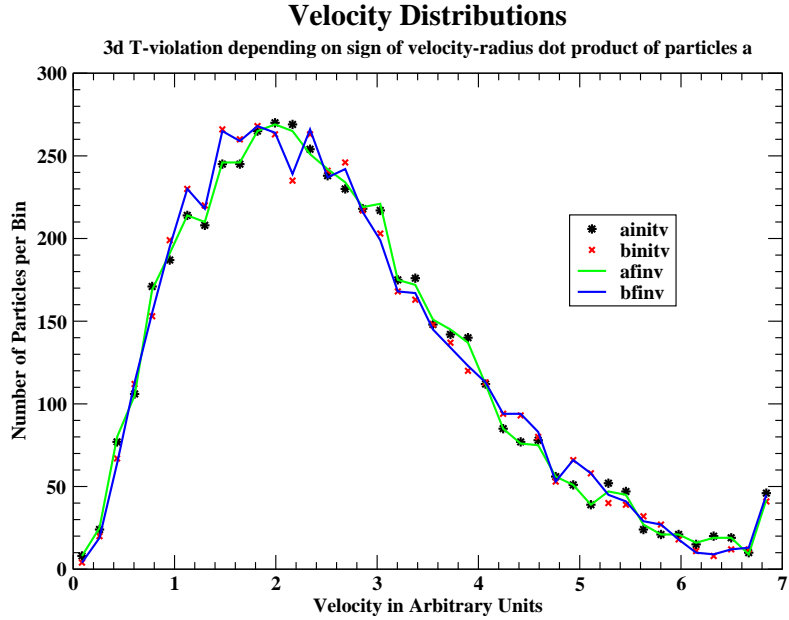}
\caption{(Color online) T-violating distributions when the scattering depends 
on  the sign of the dot product of the velocity of one species with its radial vector.
The initial (ainitv, binitv) and final (afinv, bfinv) velocity distributions for both species 
show no significant deviation from Boltzmann.  }
\label{fig:tviolva}
\end{figure}

\begin{figure}
\includegraphics[width=122mm]
{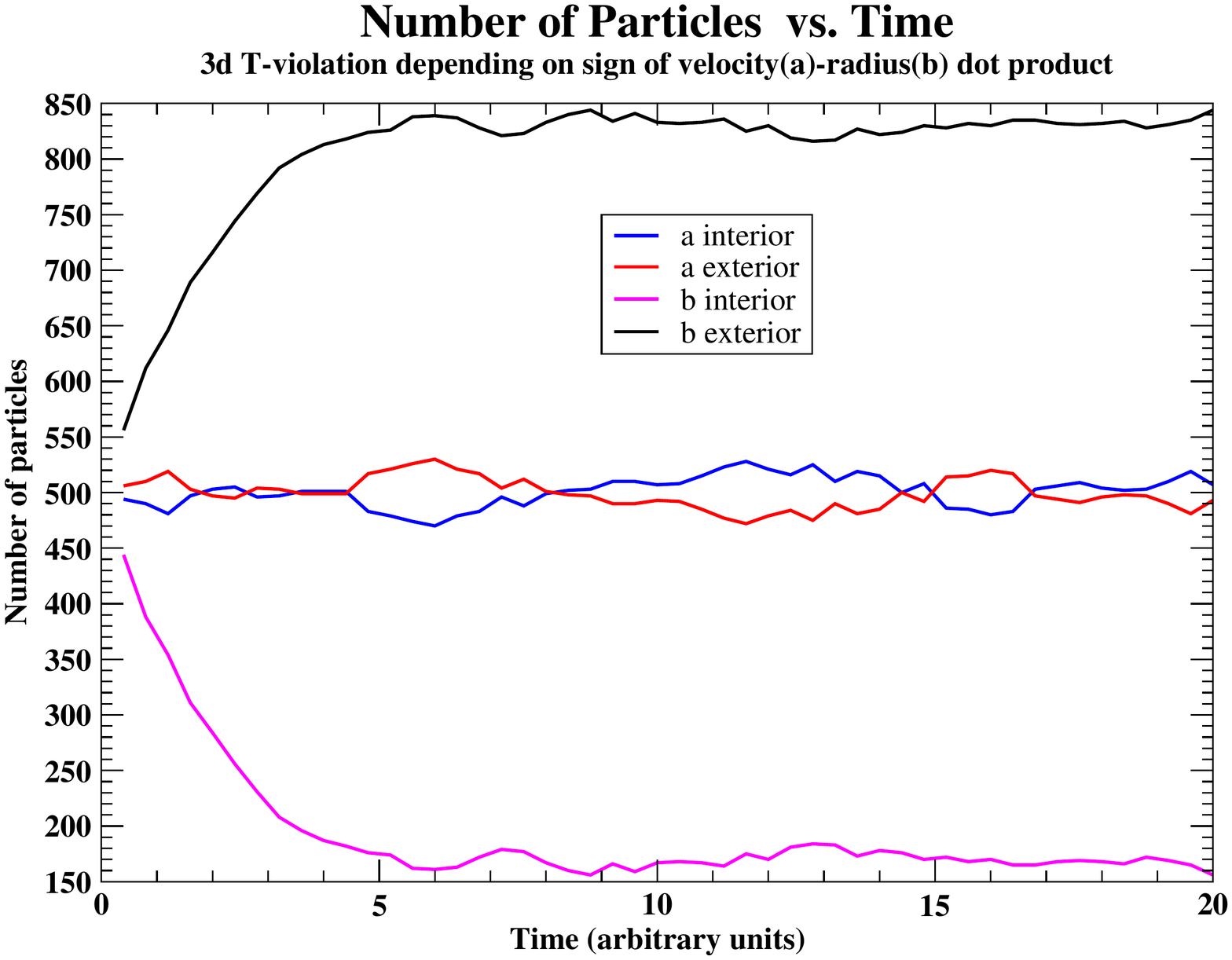}
\caption{(Color online) Time dependence of T-violating distributions when 
the scattering depends on  the sign of the dot product of the velocity of one 
species (a) with the radial vector of the other (b).}
\label{fig:tviolvarb}
\end{figure}

One might be concerned that these conclusions are limited to one dimension. 
To investigate this possibility, we consider a different closed system -- 
this time a sphere within which the two types of particles are confined, and 
subject to scattering that depends on the sign of the dot product of the 
velocity of one species with the radial vector of either the other species,
or with itself. Such a system will in general exhibit complex motions but,
as our computations show, still evolves to a final state with segregation of
the two species. See Figs.(\ref{fig:tviolra},\ref{fig:tviolva},\ref{fig:tviolvarb}).

Another concern that might be raised regarding these calculations is that we have
not explicitly derived the assumed scattering properties from a Hamiltonian. These
properties, however, have been modeled after known results in the weak interactions,
for the one-dimensional case. There the interaction is P-violating but avoids
T-violation by including C-violation, As C is not available in our model problem,
the P-violation directly translates to T-violation as noted earlier. For the
three-dimensional case, the $\; \vec{r}\cdot\vec{p} \; \; $interaction corresponds to
scattering in a fixed (T-violating) background electromagnetic field\cite{DHK}. We 
note this to confirm that our scattering matrix is derivable from a Hamiltonian; it 
does not imply that we are dealing with an open system. We conclude, therefore, 
that our examples satisfy all of the usual theoretical statistical mechanical
requirements, except for time reversal symmetry itself. 

\section{ A Known Example}

The neutral kaons are a system of hadronic particles known to be subject 
to T-violating interactions. Neutral kaons and anti-kaons decay to different 
final states, but with differing partial rates that would be identical if there 
were no T-violation. The total decay rates must be equal by the CPT theorem. 
However, if these particles are contained in a closed volume, so that the 
decay particles must recombine to form the parent particles then, starting from
an initial state containing equal numbers of each kind of particle, a net excess 
of kaons over anti-kaons (or vice versa) will develop over time due to T-violation. 
Thus, time-reversal violation produces a matter-antimatter asymmetry, albeit in
mesons, not baryons. A recently reported result from FermiLab on the
analogous process for B and anti-B mesons decaying to muons\cite{D0} finds
the number of pairs of positive muons to differ from that of negative pairs produced 
from B-anti-B meson pairs. This T-violating process could be relevant to the 
matter-antimatter asymmetry produced in the early Universe. 

\section{A Possible Application to the Early Universe}

The Universe is often assumed to begin in an organized, low entropy
state, with entropy increasing as the Universe evolves. But how
could such an organized state have come into being in the first
place? Extrapolating from the calculations above, we speculate that,
with time reversal violating forces acting for a time prior to the
onset of the Big Bang, an initially high entropy state can be
transformed into an organized, low entropy state by the time expansion
of the Universe commences. We note that recently Gurzadyan and 
Penrose\cite{GP} have used WMAP data to argue for evidence for the 
existence of a high entropy state occurring "before the Big Bang", as we 
are assuming here. 

We mention several mechanisms by which the formation of an organized,
low entropy state could occur. First we consider a physical system
with a  fourth spatial dimension and imagine that the 3-dimensional
world (3-brane) of our  visible Universe is the surface of a 4-sphere
in that larger space. Starting from a well mixed (high entropy)
distribution of, say, matter and antimatter over a restricted range
in the 4-radial direction, and with the time reversal violating forces
acting along  that line, we infer on the basis of the calculations
presented above (recall Fig.\ref{fig:tviolmax}) that the system will separate
into parts with matter on one 3-brane and antimatter on a distinct
3-brane. In other words, the system evolves from what would normally
be described as a high entropy (well-mixed) state to one of lower
entropy and high density on our 3-brane, which state would then expand
our 3-brane following standard cosmology.

Alternatively, and more conventionally, we can imagine that in every
small volume of the early Universe, {\em elastic} T-violating scattering
in the radial direction separates matter and antimatter, again as in our
1-dimensional example calculation. Now as the  Universe expands, these
separated clumps may remain separated rather than remix. This leads to a
picture where there need not be a matter asymmetry in the Universe, but
where the matter-antimatter annihilation radiation is suppressed by the
growth in size of the distinct regions of matter and of antimatter. 

Sakharov\cite{Sakh} has discussed the development of excess baryon number 
in the evolution of the early Universe. His analysis requires three elements: 
baryon number violation, a non-equilibrium state, and time reversal violation.
The first is obvious; regarding the second, we raise the question of how a
state of thermodynamic equilibrium should be defined in the presence of P- 
and T-violating dynamics and the last we view as the essential element, noting
that an excess of baryons over anti-baryons represents a more ordered state than
does equality. Thus, this too may be viewed as an example of the development
of a more ordered state (by conventional definition) due to a violation of
time reversal invariance.  

Thus, we observe that time-reversal violation may obviate two of the conditions 
assumed in Sakharov's discussion\cite{Sakh} of the matter/antimatter problem: 
Our 5-dimensional system example is in what would normally be considered 
equilibrium, but nonetheless develops structure and additionally separates 
matter and antimatter, absent even a baryon-number violating interaction. 
The only thing that we have maintained is the requirement of T (equivalently 
CP) violation.

Another mechanism\cite{pat} describes a possible origin of our Universe as 
the white hole arising from 
an Einstein-Rosen bridge connected to a Weyl metric black hole in another 
Universe. Note that at the event horizon, the space and time coordinates 
are interchanged since the metric signature for each changes sign; the 
continuation of what would have been space in a Schwarzchild black hole 
metric is converted (essentially by the equivalent of coordinate inversion in a 
sphere) to an infinite time future, and correspondingly time into space. 

Two significant points follow: Since all of the material being transported from 
the original Universe to ours crosses the same radial coordinate value, all of 
that material appears to have originated in our Universe at a common initial 
time, at a high initial density and temperature, corresponding to a low entropy 
state. In fact, since the proper radial velocity becomes infinite, the motion of the
particles entering the (new) Universe is aligned in flow (up to quantum
fluctuations not considered here) so that the initial state must be viewed as
maximally organized, i.e., it has minimum entropy.  Secondly, since time reversal
is violated, the excess of baryons over anti-baryons in our Universe can be due to 
an initial condition, namely the in-falling baryon content.  Thus only one of Sakharov's 
requirements, T-violation, remains essential. 

\section{Conclusion}

The ideas discussed here have been preceded by more general considerations of 
Dolgov and Zeldovich\cite{DZ} and by Cohen and Kaplan\cite{CK}, who also explored 
the limits of the conventional assumptions. Additionally, Bertolami et al.,\cite{Bert} have 
shown how CPT violation in the expanding Universe defines a chemical potential 
so that a baryon asymmetry may develop while in a state of conventional thermal
equilibrium. In fact, in this scenario, the asymmetry is so large that it must
be diluted by electroweak sphaleron effects to match observation. In any event,
it also produces an organized state from a disorganized one, as we suggest here. 

We have illustrated how more ordered states may evolve from less ordered ones
when time-reversal-violating interactions are active. Thus, with the conventional
definition of entropy, time-reversal-violating interactions can produce reductions
in entropy in closed systems. This implies that the ordered state of the early
Universe and the observed baryon asymmetry may have been produced by 
time-reversal-violating interactions. We raise the question of whether or not the 
conventional definition of entropy requires revision to account for the inaccessibility 
of some phase space states in the absence of time-reversal-invariance. 

\section{Acknowledgments}
This work was carried out in part under the auspices of the National Nuclear Security
Administration of the U.S. Department of Energy at Los Alamos National Laboratory
under Contract No. DE-AC52-06NA25396. We have benefitted from discussions on 
this topic with Harvey Rose, Cynthia Olsen Reichardt, Charles Reichardt, Daniel Holz 
and Salman Habib.

\end{document}